\documentclass [10pt]{article}

\usepackage{amsmath}

\usepackage{slashed}

\usepackage{hyperref}
\usepackage{geometry}

\usepackage{graphicx}
\usepackage{amssymb}
\usepackage{epstopdf}

\usepackage{graphicx}
\usepackage[usenames]{color}
\usepackage{subfigure}
\usepackage{setspace}
\usepackage{slashed}
\usepackage{multirow}
\usepackage{ulem}

\usepackage[hang,small]{caption}

\usepackage{geometry}
   \usepackage{amsmath}   
        \usepackage{amssymb}   
    
    \newcommand{\ignore}[1]{}
 \newcommand{\ee}{\end{equation}}

\def\ba#1\ea{\begin{align}#1\end{align}}
\newcommand{\bit}{\begin{itemize}}
\newcommand{\eit}{\end{itemize}}

\newcommand{\nn}{\nonumber} \renewcommand{\bf}{\textbf}
\newcommand{\ra}{\rightarrow}

\renewcommand{\a}{\alpha} 
\newcommand{\e}{\mathrm{e}} 
  \newcommand{\g}{\gamma}

\newcommand{\ii}{{i\mkern1mu}}

\usepackage{fancyhdr}
\usepackage{titlesec,titletoc}
  \titleformat{\section}{\Large\sf\bfseries}{\thesection}{1em}{}
  \titleformat{\subsection}{\large\sf\bfseries}{\thesubsection}{1em}{}

\title{\sf\bfseries \ntitle}
\author{
    Sumeet Dagaonkar\footnote{sumeetkd@iitk.ac.in}~ \\ 
    \it{Department of Physics, Indian Institute of Technology, Kanpur 208016, India}\\
}

\date{}

 \newcommand{\pghdr}{\footnotesize {S. Dagaonkar} -- Compton scattering  \dots }
\newcommand{\ntitle}{Compton scattering in the Endpoint Model}

\begin{document}

\maketitle
\begin{abstract}
{
We use the Endpoint model for exclusive hadronic processes to study Compton scattering of the
proton. The parameters of the Endpoint model are fixed using the data for $F_1$ and the ratio
of Pauli and Dirac form factors ($F_2/F_1$) and then used to get numerical predictions for the
differential scattering cross section. We studied the Compton scattering at fixed $\theta_{CM}$ in
the $s \sim t \gg \Lambda_{QCD}$ limit and at fixed $s$  much larger than $t$ limit. We
observed that the calculations in the Endpoint Model give a good fit with experimental data in both
regions.
}   
\end{abstract}

Though we have a well understood QCD Lagrangian, predicting processes involving hadrons is a
difficult task. The interaction of a high energy probe with quarks or gluons in a hadron
requires us to understand physics which is non-perturbative. While in processes like deep inelastic
scattering we are able to successfully use factorization - to separate the non-perturbative
part into a parton distribution function while the rest could be calculated perturbatively, such
simplifications are understood to be much more difficult in the case of exclusive processes \cite{Isgur:1984jm}
.
Theoretical models aimed at explaining such processes have been around for four decades now and the ideas can
be spilt into two major camps - methods involving hard gluon exchanges within the constituents
(short distance model) and methods without hard exchanges (soft or Feynman mechanism). The Endpoint
Model(EP) used in this paper combines the idea of soft mechanism with a model of hadron wavefunction which
constrains the transverse momenta of confined quarks.

The exclusive process of interest in this paper is the Real Compton scattering ($p\gamma \ra
p\gamma$).
The first measurements for Compton scattering were made at Cornell \cite{Shupe:1979vg}, where the
differential cross section $d\sigma/dt$ was measured and found to show a scaling of $1/s^6$. However more recent
measurements at JLab \cite{Danagoulian:2007gs} have shown that the scaling goes more like $1/s^{8.0
\pm 0.2}$. In recent years the experiments using polarization transfer \cite{Fanelli:2015eoa} have also given measurements
of transverse polarization transfer $K_{LS}$ and longitudinal polarization transfer $K_{LL}$.

The first theoretical predictions for the scaling behaviour of Compton scattering appeared in
\cite{Brodsky:1973kr,Matveev:1973ra}. They predicted that $ d\sigma/dt|_{\mathrm{fixed\, t}} \propto
1/s^{6} f(t/s)$ using simple constituent counting ideas. 
Recent calculations in perturbative QCD (short distance model)  \cite{Brooks:2000nb,Thomson:2006ny}
using this formalism give predictions a order lower than the experimental data. However, it is
understood that the perturbative calculations are only applicable at asymptotically high energies
not explored at existing experimental facilities.
The soft mechanism was used by Diehl et al.\cite{Diehl:1998kh} in calculations involving
generalized parton distribution functions (GPD), while Miller \cite{Miller:2004rc} calculated the
handbag diagram in the constituent quark model (CQM).  The former work was shown to be equivalent to
a sum of overlap of light cone wave functions for all Fock states. For the leading Fock state, the
pole structure leads to a similar endpoint dominance as obtained in our model. While the GPD based
analysis agrees with some features of the data, the scaling behaviour is not consistent with
the latest data.
Work by Kivel and Vanderhaeghen \cite{Kivel:2013sya,Kivel:2015vwa} on Compton scattering unifies 
the short distance and the soft mechanism using Soft collinear effective theory.

The latest results on polarization transfer measurements \cite{Fanelli:2015eoa} show that, while the
$K_{LS}$ agrees well with the results of pQCD\cite{Thomson:2006ny},
GPD's \cite{Huang:2001ej}, CQM \cite{Miller:2004rc} and SCET \cite{Kivel:2015vwa}, the
$K_{LL}$ measurements have been unexpectedly larger and do not
agree with any of the theoretical predictions.

The Endpoint Model \cite{Dagaonkar:2014yea,Dagaonkar:2015laa} applies to all exclusive hadronic
processes and reproduces the quark counting rules at high energies.
In the model, the dominant contributions 
involve struck quarks carrying a large fraction of the hadron's momenta.
The scaling is now completely dependent on the endpoint behaviour of the light cone wavefunctions.
It is then possible to obtain the functional form of the wavefunction near the endpoint.
After extracting the wavefunction of the proton from the $F_1$ data, the authors successfully used
the wavefunction to understand the scaling behaviour of $pp$ scattering and the ratio of the Pauli and
Dirac form factors ($F_2(Q^2)/F_1(Q^2)$) of the proton. These results motivated the author to
attack the Compton scattering problem using the Endpoint Model.

After introducing the Endpoint Model and setting up the frame work, we will show in Section
\ref{sec:1} that the EP calculation for $d\sigma / dt$ obeys scaling laws of
\cite{Brodsky:1973kr,Matveev:1973ra} at large $s$ in the $s \sim t \gg \Lambda_{QCD}$ limit and also a
scaling of $1/t^4$ in the fixed  $s$ much larger than $t$ region. A detailed numerical calculation in Section
\ref{sec:2} will help us determine the range of $s$ for which we may expect the scaling behaviour to
set in and we will also extend the model's prediction into a low $Q^2$ region to compare with data.
At asymptotic energies, we expect that pQCD contributions may dominate. However as seen in the
current analysis, a soft mechanism like the Endpoint model can be used to understand data
which lies within experimental reach.

\section{Compton scattering using the Endpoint model}
\label{sec:1}

\begin{figure}[h]
\centering
\includegraphics[scale = .45]{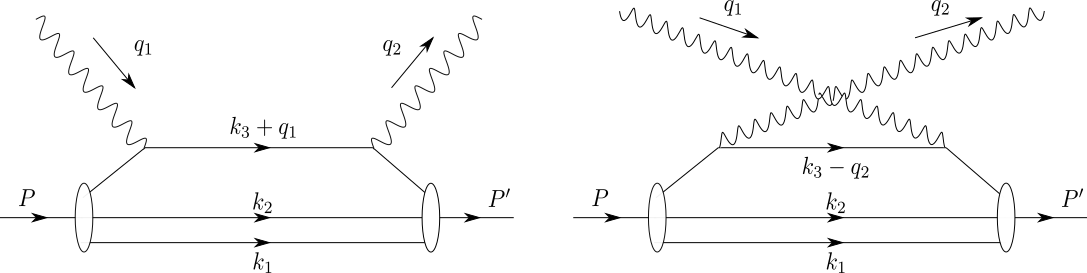}
\caption{Two diagrams from Compton scattering of proton in the Endpoint Model}
\label{fig:cs2diag}
\end{figure}

The diagrams allowed for Compton scattering under the Endpoint Model are given in Fig.
\ref{fig:cs2diag}. 
It can be noticed that the interaction between the struck quark and the photon mirrors the diagrams
of the Compton scattering with electrons.

\subsection{Kinematics}
In the above diagrams, the incoming proton is understood to be deflected by $q^{\mu} =
(0,Q,0,0)$, where $q^{\mu} = q_{1}^{\mu} - q_{2}^{\mu}$. This allows us to use the same frame and
kinematics for the proton, as was used for the analysis of Dirac and Pauli form factors
\cite{Dagaonkar:2014yea,Dagaonkar:2015laa} with
$q=(0,Q,0,0), 
P=(\sqrt{Q^{2}/2+M_{P}^{2}},-Q/2,0,Q/2),
P'=(\sqrt{Q^{2}/2+M_{P}^{2}},Q/2,0,Q/2)$.
We can choose $q_1,q_2$ appropriately so that $\theta_{cm} \in [64^{\circ},  130^{\circ}]$, which is the
range of the data obtained at Jlab \cite{Danagoulian:2007gs}.
For $\theta_{cm} \approx 90^{\circ}$,
$q_1=\left(Q/\sqrt{2},Q/2,0,-Q/2\right),
q_2=\left(Q/\sqrt{2},-Q/2,0,-Q/2\right).$

Let us also define the various quark momenta that will be useful in our calculation, starting with a
basis for transverse momenta: $ y^{\mu} =(0, 0, 1,0) = y'$  such that $ \hat P\cdot y=\hat
P'\cdot 
 y'=0$ and   $n^{\mu} = (1/\sqrt{2}) (0, -1, 0, -1)$ such that $  \hat P\cdot n= 0$ and 
$n'^{\mu} = (1/\sqrt{2}) (0, 1, 0,-1)$ such that 
 $\hat P'\cdot n'=0.$
Here $\hat P = \left(0,-1/\sqrt{2},0,1/\sqrt{2}\right)$ and $\hat P' =
\left(0,1/\sqrt{2},0,1/\sqrt{2}\right)$ are the unit vectors along
the direction of propagation of the incoming photon and incoming proton respectively.
The four momenta of the quarks are then given by,
\ba
k_{i}^{\mu} &=
\left(k_{i}^{0},-x_{i}\frac{Q}{2}-\frac{k_{in}}{\sqrt{2}},k_{iy},x_{i}\frac{Q}{2}-\frac{k_{in}}{\sqrt{2}}\right)
\nn\\
k_{i}^{'\mu} &=
\left(k_{i}^{'0},x'_{i}\frac{Q}{2}+\frac{k'_{in}}{\sqrt{2}},k'_{iy},x'_{i}\frac{Q}{2}-\frac{k'_{in}}{\sqrt{2}}\right)
\label{eq:qmomentum}
\ea
\subsection{Endpoint Model Calculation}
The amplitude for the process can be written as
\ba
\label{eq:amp}
\ii\mathcal{M} = \int \prod\limits_{i} \frac{d^4 k_i}{(2\pi)^4} \frac{d^4 k'_i}{(2\pi)^4}
(2\pi)^4\delta(k_1+k_2+k_3-P) (2\pi)^4\delta(k'_1+k'_2+k'_3-P') \\ \nn
\epsilon^{*\mu}(q_{2})\epsilon^{\nu}(q_{1})\left[ \overline{\Psi'}_{\alpha'\beta'\gamma'}(k^{'}_{i}) \times
\mathcal{M}_{\alpha'\beta'\gamma'\alpha\beta\gamma}^{\mu\nu}\times \Psi_{\alpha\beta\gamma}(k_{i})
\right],
\ea
where $\Psi_{\alpha\beta\gamma}$ refer to 3 quark Bethe-Salpeter wavefunction, the indices
$\alpha,\beta,\gamma$ refer to the $u,u,d$ carrying momentum $k_1,k_2,k_3$ respectively. The primed
quantities refer to the outgoing proton.

The $\mathcal{M}^{\mu\nu}$ in the above expression is taken as,
\begin{eqnarray} 
\mathcal{M}_{\alpha'\beta'\gamma'\alpha\beta\gamma}^{\mu\nu}&=&
\left[(-\ii e_{u}\gamma^{\mu}) \frac{\ii (\slashed{k}_1+\slashed{q}_1+m_q)}{(k_1+q_1)^2-m_q^2}(-\ii
e_{u}\gamma^{\nu}) + (-\ii e_{u}\gamma^{\nu}) \frac{\ii (\slashed{k}_1-\slashed{q}_2+m_q)}{(k_1-q_2)^2-m_q^2}(-\ii
e_{u}\gamma^{\mu})  \right]_{\alpha^{'}\alpha}\nn \\ 
&& (2\pi)^{12}\delta^{4}(k_{1}+q-k'_{1})\ii(\lambda\slashed{k}_{2}-m_{2})_{\beta^{'}\beta}
\delta^{4}(k_{2}-k'_{2})\ii(\lambda\slashed{k}_{3}-m_{3})_{\gamma^{'}\gamma}\delta^{4}(k_{3}-k'_{3})  \nn \\
&+&\left[(-\ii e_{u}\gamma^{\mu}) \frac{\ii (\slashed{k}_2+\slashed{q}_1+m_q)}{(k_2+q_1)^2-m_q^2}(-\ii
e_{u}\gamma^{\nu}) + (-\ii e_{u}\gamma^{\nu}) \frac{\ii (\slashed{k}_2-\slashed{q}_2+m_q)}{(k_2-q_2)^2-m_q^2}(-\ii
e_{u}\gamma^{\mu})  \right]_{\beta^{'}\beta}\nn \\  & & (2\pi)^{12}\delta^{4}(k_{2}+q-k'_{2})\ii(\lambda\slashed{k}_{1}-m_{1})_{\a^{'}\a}\delta^{4}(k_{1}-k'_{1})
\ii(\lambda\slashed{k}_{3}-m_{3})_{\g^{'}\g}\delta^{4}(k_{3}-k'_{3})  \nn\\
&+&\left[(-\ii e_{d}\gamma^{\mu}) \frac{\ii (\slashed{k}_3+\slashed{q}_1+m_q)}{(k_3+q_1)^2-m_q^2}(-\ii
e_{d}\gamma^{\nu}) + (-\ii e_{d}\gamma^{\nu}) \frac{\ii (\slashed{k}_3-\slashed{q}_2+m_q)}{(k_3-q_2)^2-m_q^2}(-\ii
e_{d}\gamma^{\mu})  \right]_{\gamma^{'}\gamma}\nn \\  &&
(2\pi)^{12}\delta^{4}(k_{3}+q-k'_{3})\ii(\lambda\slashed{k}_{1}-m_{1})_{\alpha^{'}\alpha}\delta^{4}(k_{1}-k'_{1})\ii(\lambda\slashed{k}_{2}-m_{2})_{\beta^{'}\beta}\delta^{4}(k_{2}-k'_{2}),
\label{eq:m}
\end{eqnarray}
where we have taken into account both diagrams in Fig. [\ref{fig:cs2diag}] and the three terms represent
the photon's interactions with u,u,d quarks respectively.

We would like to integrate over the $k_i^-, k_i^{'-}$ momenta in the Eq. \ref{eq:m}  so as to replace the Bethe Salpter
wavefunctions by Light cone wavefunctions using the approximations developed in \cite{Brodsky:1984vp}.
The integrand has $k_i^-, k_i^{'-}$ dependence 
due to the propagators associated with the Bethe Salpeter wavefunction and
from the spectator quarks. The spectator quarks interact through soft gluons and 
behave like a effective diquark propagator. Its form will require us to do an detailed analysis of the
physics in this non-perturbative system. As a starting point however, we use a simple model consisting of two non-interacting
quarks given by $(\lambda \slashed{k}_2 - m)(\lambda \slashed{k}_3 - m)$,
where $\lambda$ may be a scalar function of the spectator quark momentum ($k_2,k_3$).
The complete expression for $\mathcal{M}^{\mu\nu}$ is assumed to be dominated by a region where the quarks
are on-shell which allows us to make the substitution $\kappa_i^- = (k^0 - x_i
Q/\sqrt{2})(P^0+Q/\sqrt{2})= (m_i^2+\vec{k}^2_{\perp \, i})/(k^{0} + x_i Q/\sqrt{2})$. In this substitution, we have taken into
account the energy scale dependence of the mass which causes the effective mass to be $m^2_i\sim
\Lambda^2$ for the spectator quarks and $m^2_i \sim$ few MeV for the struck quark. 
Momenta for each of the quarks is conserved independently and as per the definition of
$i\mathcal{M}$, a factor of $\delta^{4}(P+q_1 - q_2 -P')$ has to be dropped in the above expression.

Under these approximations, the amplitude \ref{eq:amp} becomes
\begin{eqnarray}
\ii\mathcal{M} &=& \int \prod\limits_{i} dx_{i}d\vec{k}_{\perp i} 
dx^{'}_{i}d\vec{k}'_{\perp i}  \delta(x_{1}+x_{2}+x_{3}-1)\delta^{2}(k_{\perp 1}+k_{\perp 2}+k_{\perp
3})
\delta(x'_{1}+x'_{2}+x'_{3}-1)\nn \\ & &  \delta^{2}(k'_{\perp 1}+k'_{\perp 2}+k'_{\perp 3})
\epsilon^{*\mu}(q_{2})\epsilon^{\nu}(q_{1})\left[ \overline{Y^{\prime}}_{\alpha^{'}\beta^{'}\gamma^{'}}(x'_{i},\vec{k}'_{\perp
i})
 \times
\mathcal{M}_{\alpha'\beta'\gamma'\alpha\beta\gamma}^{\mu\nu}\times Y_{\alpha\beta\gamma}(x_{i},\vec{k}_{\perp i})
\right].
\label{eq:bjs}
\end{eqnarray}
The light cone wave function for the proton $Y(k_i)$ at leading twist and leading power of large $P$ is
\cite{Ioffe,Avdeenko},
 \begin{equation}      Y_{\alpha\beta\gamma}(k_{i},P) = \frac{f_{N}}{16\sqrt{2}N_{c}}\{ 
     (\slashed{P}C)_{\alpha\beta}(\gamma_{5}N)_{\gamma}\mathcal{V} + (\slashed{P}\gamma_{5}C)_{\alpha\beta}N_{\gamma}\mathcal{A} + \ii (\sigma_{\mu\nu}P^{\nu}C)_{\alpha\beta}
 (\gamma^{\mu}\gamma_{5}N)_{\gamma}\mathcal{T}\}.\label{eq:lipwavef}
\end{equation}
 Here $\mathcal{V,A,T}$ are scalar wavefunctions of the quark momenta,
         $N$ is the proton spinor,
        $N_{c}$ the number of colors,
        $C$ the charge conjugation operator, 
        $\sigma_{\mu\nu}= \frac{\ii}{2}[\gamma_{\mu},\gamma_{\nu}]$,
and $f_{N}$ is a normalization. 
The functional dependence for the scalar functions near the endpoint region of the $x_i$, the
momentum fraction of the struck quark, was
obtained in \cite{Dagaonkar:2014yea} by matching the EP calculation with the experimental scaling behaviour of
$F_1$ of the proton. We will carry over that form in this paper
\ba
\mathcal{V} = v (1-x_i)\e^{- k_{T}^{2}/\Lambda^{2}};\hspace{5mm}\mathcal{A} = a (1-x_i)\e^{-
k_{T}^{2}/\Lambda^{2}};\hspace{5mm}\mathcal{T} = t (1-x_i)\e^{- k_{T}^{2}/\Lambda^{2}}
\label{eq:wf}.
\ea
The $\vec{k}_{T}$ represents the
transverse momenta of the quark which is suppressed by an exponential function in the above form and
is understood to be cut off sharply for $|k_T| > \Lambda_{QCD}$.

\subsection{Scaling in Endpoint Model}
\label{sec:scaling}

Before presenting the endpoint model's prediction for Compton scattering, we explicitly
evaluate a part of the entire expression to extract the scaling behaviour to be expected for 
fixed $\theta_{CM}$ and fixed $s$ cases.

Let us concentrate on the diagram shown in Fig. \ref{fig:cs2diag}, in which $d$ quark is struck. 
The delta functions in the last term of Eq. \ref{eq:m} and 
Eq. \ref{eq:bjs} imply,
$  x_{1} = 1 - x_{2} - x_{3}; x'_{1} = 1 - x'_{2} - x'_{3};   k_{1n} = -k_{2n} -
k_{3n};  k_{1y} = -k_{2y} - k_{3y} ;   
  k'_{1n} = -k'_{2n} - k'_{3n};  k'_{1y} = -k'_{2y} - k'_{3y} ;  
 k_{2y} = k'_{2y}; k_{3y} = k'_{3y} ;\, x'_{2}= x_{2}; x'_{3}= x_{3}; 
 k_{3n}=Q(1-x'_{3})/\sqrt{2} ;\, k'_{3n}=Q(1-x_{3})/\sqrt{2}; 
 k_{2n}=Q(-x'_{2})/\sqrt{2} ;\, k'_{2n}= Q(-x_{2})/\sqrt{2}. 
$

Integrating over the delta functions leads to a factor of $1/Q^2$. Using only the first term of the
wavefunction Eq. \ref{eq:lipwavef}, the amplitude is obtained as,
\begin{eqnarray}
i\mathcal{M} &=& \int dx_1 dx_2 dk_{1y} dk_{2y} \frac{1}{Q^2}\hspace{5pt}
\epsilon^{*\mu}(q_{2})\epsilon^{\nu}(q_1)\bigg[[(C^{-1}\slashed{P'})_{\alpha'\beta'} (\overline{N}\gamma_5)_{\gamma'}\mathcal{V}^*]\nn \\ 
&& \left[(-\ii e_{d}\gamma^{\mu}) \frac{\ii (\slashed{k_3}+\slashed{q_1}+m_q)}{(k_3+q_1)^2-m_q^2}(-\ii
e_{d}\gamma^{\nu}) + (-\ii e_{d}\gamma^{\nu}) \frac{\ii (\slashed{k_3}-\slashed{q_2}+m_q)}{(k_3-q_2)^2-m_q^2}(-\ii
e_{d}\gamma^{\mu})  \right]_{\gamma^{'}\gamma}\nn \\  &&
\ii(\lambda\slashed{k}_{1}-m_{1})_{\alpha^{'}\alpha}\ii(\lambda\slashed{k}_{2}-m_{2})_{\beta^{'}\beta}
[(\slashed{P}C)_{\alpha\beta}(\gamma_{5}N)_{\gamma}\mathcal{V}]+\cdots \bigg]
\end{eqnarray}
The experimentally measured quantity is the unpolarized differential cross section $d\sigma/dt= 1/16\pi(s-m_p^2)^2
1/4\sum|M|^2$, (the integrations in the complex conjugate are over the hatted variables) 
\begin{eqnarray}
\frac{d\sigma}{dt} &=& \frac{1}{16\pi(s-m_p^2)^2}\frac{1}{4}\int dx_1 dx_2 dk_{1y} dk_{2y} \frac{1}{Q^2} \int d\hat{x}_1 d\hat{x}_2 d\hat{k}_{1y} d\hat{k}_{2y} \frac{1}{Q^2}\nn \\
&&\bigg[\mathrm{Tr}[(C^{-1}(\slashed{P'})(\lambda\slashed{k}_{2}-m_{2})(\slashed{P}C)^\intercal(\lambda\slashed{k}_{1}-m_{1})^\intercal]
\mathrm{Tr}[(C^{-1}(\slashed{P'})(\lambda\slashed{\hat{k}}_{2}-m_{2})(\slashed{P}C)^\intercal(\lambda\slashed{\hat{k}}_{1}-m_{1})^\intercal]^*
\nn \\
&& \mathrm{Tr}\bigg[(\slashed{P'}+m_p)\gamma_5[ \frac{\gamma^{\mu}
(\slashed{k}_3+\slashed{q}_1+m_q)\gamma^{\nu}}{(k_3+q_1)^2-m_q^2} + \frac{ \gamma^{\nu}(\slashed{k}_3-\slashed{q}_2+m_q)\gamma^{\mu}}{(k_3-q_2)^2-m_q^2}
]\gamma_5  (\slashed{P}+m_p)\gamma_5 \nn \\ && 
[\frac{\gamma^{\nu'}
(\slashed{\hat{k}}_3+\slashed{q}_1+m_q)\gamma^{\mu'}}{(\hat{k}_3+q_1)^2-m_q^2} +  \frac{
\gamma^{\mu'}(\slashed{\hat{k}}_3-\slashed{q}_2+m_q)\gamma^{\nu'}}{(\hat{k}_3-q_2)^2-m_q^2}]\gamma_5\bigg]
e^4_{d}\mathcal{V}^*(k'_i)\mathcal{V}(k_i)\mathcal{V}^*(\hat{k}'_i)\mathcal{V}(\hat{k}_i)\nn +
\cdots \bigg] \\
&& \sum\limits_{polarization}
\epsilon^*_{\mu}(q_2)\epsilon_{\mu'}(q_2) \sum\limits_{polarization}
\epsilon^*_{\nu}(q_1)\epsilon_{\nu'}(q_1).
\label{eq:dsdt}
\end{eqnarray}
We can integrate over the variables after plugging in the wavefunction from Eq. \ref{eq:wf}.
Our calculation shows scaling behaviour in two limits, for $s \sim t \gg \Lambda_{QCD}$ 
and for fixed $s$ much larger than $t$. 
In the $s \sim t \gg \Lambda_{QCD}$ limit, the leading order contributions give, 
\ba
\frac{d\sigma}{dt} &\sim \int dx_1 dx_2 dk_{1y} dk_{2y} \int d\hat{x}_1 d\hat{x}_2 d\hat{k}_{1y}
d\hat{k}_{2y} \frac{1}{16\pi(s-m_p^2)^2}\left(\frac{1}{Q^2}\right)^2((P\cdot P')(k_1\cdot
k_2)+\dots)\times\nn \\ & \hspace{15pt}((P\cdot P')(\hat{k}_1\cdot \hat{k}_2)+\dots) 
\frac{((k_3\cdot P)(\hat{k}_{3}\cdot P')+\dots)}{(k_3\cdot q_1)(\hat{k}_3 \cdot
q_2)}(1-x_3)^2 (1-\hat{x}_{3})^2 + \cdots \nn \\ &
\sim
\frac{1}{s^2}\left(\frac{1}{Q^2}\right)^2(Q^2)^2\frac{(Q^2)^2}{(Q^2)^2}\frac{1}{Q^4}\frac{1}{Q^4}
\sim \frac{1}{s^2}\times\frac{1}{s^2}\times\frac{1}{s^2} \sim \frac{1}{s^6}.
\ea 
Thus in the large $s$ limit, we can see that we obtain a scaling behavior of $1/s^6$, as expected
from the quark counting rules.

In order to analyse the differential cross section for fixed $s$  when $s > t$, we have to alter the
photon momenta defined specifically for $\theta_{CM} = 90^{\circ}$ above and instead use $
q_1=\left(Q/\sqrt{2}, Q/2,0,f(s,Q)\right)$, $ q_2=\left(Q/\sqrt{2},-Q/2,0,f(s,Q)\right)$.
The definition of $s = (P+q_1)^2$ can be used to find the functional form of $f(s,t)$. To the
leading order in $s$, it can be shown that $f(s,Q) \sim \pm s/Q$.
In the $s > t $ limit, the leading order contributions are now,
\ba
\frac{d\sigma}{dt} &\sim \int dx_1 dx_2 dk_{1y} dk_{2y} \int d\hat{x}_1 d\hat{x}_2 d\hat{k}_{1y}
d\hat{k}_{2y} \frac{1}{16\pi(s-m_p^2)^2}\left(\frac{1}{Q^2}\right)^2((P\cdot P')(k_1\cdot
k_2)+\dots)\times\nn \\ & \hspace{15pt}((P\cdot P')(\hat{k}_1\cdot \hat{k}_2)+\dots) 
\frac{((q_2\cdot P)(q_1\cdot P')+\dots)}{(k_3\cdot q_1)(\hat{k}_3 \cdot
q_2)}(1-x_3)^2 (1-\hat{x}_{3})^2 + \cdots \nn \\ &
\sim
\frac{1}{s^2}\left(\frac{1}{Q^2}\right)^2(Q^2)^2\frac{s^2}{s^2}\frac{1}{Q^4}\frac{1}{Q^4} \sim
\frac{1}{Q^4}\times\frac{1}{Q^4} \sim \frac{1}{Q^8} \sim \frac{1}{t^4}
\ea

\section{Comparing Compton scattering in Endpoint Model with experimental data}
\label{sec:2}

The full prediction of the Endpoint Model for Compton scattering involves substituting the full
expressions Eq.  \ref{eq:m}, \ref{eq:lipwavef} into the expression Eq. \ref{eq:dsdt}. The expression
involves multiple traces over gamma matrices which were handled by the Mathematica package FEYNCALC
\cite{Mertig:1990an}. The resulting expression contains thousands of terms for each combination of
wavefunction $\mathcal{V,A,T}$.
Due to the large number of terms, analytic evaluation would be cumbersome and it is dealt with
using a Monte Carlo routine for integration (VEGAS \cite{Lepage:1977sw}).

In the previous work on Endpoint Model \cite{Dagaonkar:2014yea}
, the authors concentrated on explaining the scaling behaviour
of exclusive hadronic processes using a functional form of the wavefunction. In the current paper, we
extended the above work by using $\chi^2$ minimization to extract the free parameters of the model.
This would be essential when comparing the magnitude of the prediction of the Compton scattering in
EP with data. Using the data for
$F_1$\cite{Sill:1992qw} and $F_2/F_1$\cite{Puckett:2010ac,Puckett:2011xg}(at $Q^2 \gtrsim 5.5
\mathrm{\,GeV}^2$) and the EP prediction in
\cite{Dagaonkar:2015laa}
, the minimization gives the values for the constants $v,a,t$ from Eq.
\ref{eq:wf}, mass of the quark $m$ and the factor $\lambda$ for the model of spectator quarks.
 The constants obtained in the above minimization will carry over
to all the processes that EP may be applied to.

\begin{figure}[h]
\includegraphics[width=3.8in,keepaspectratio]{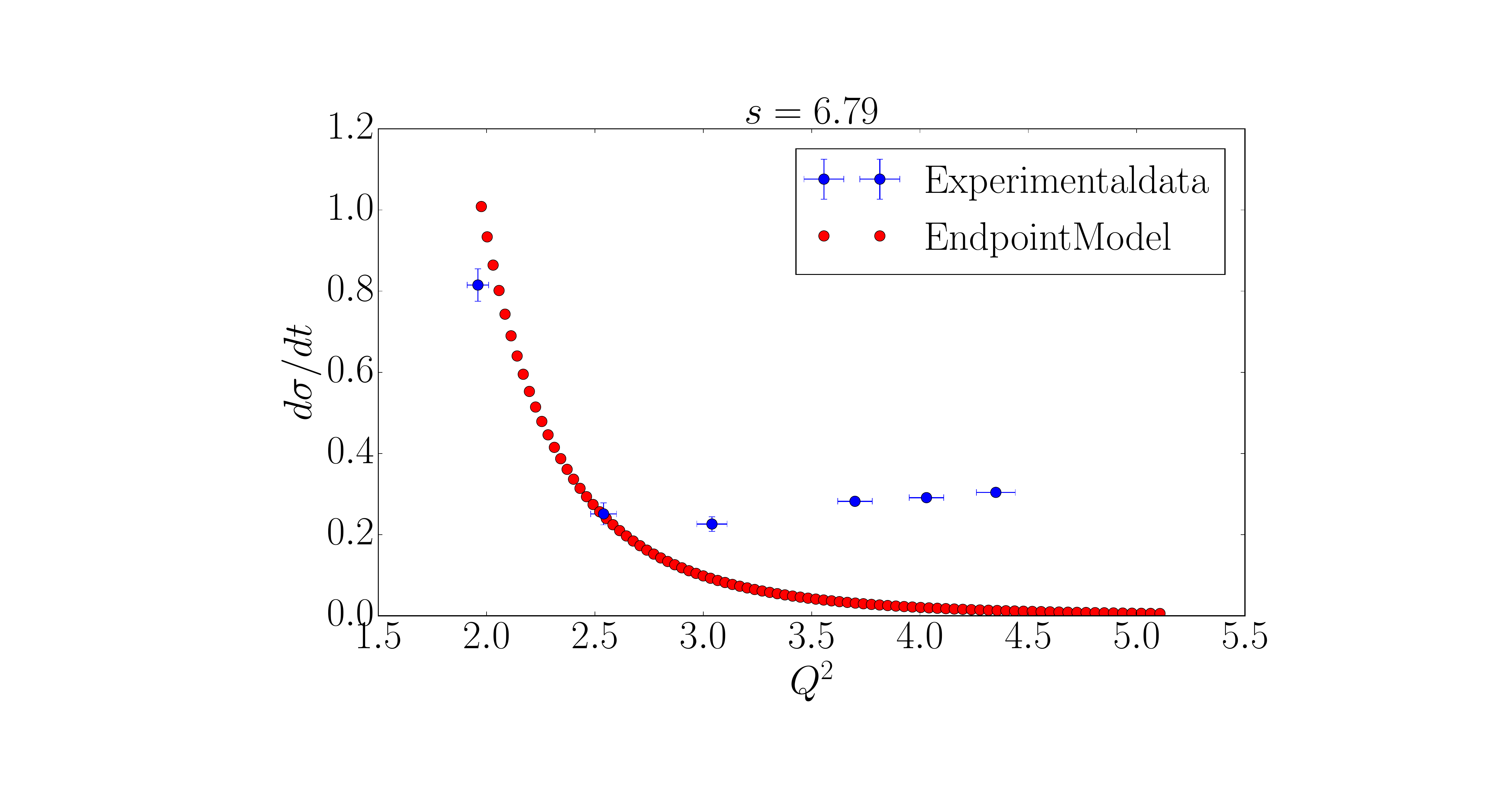}\\
\includegraphics[width=3.8in,keepaspectratio]{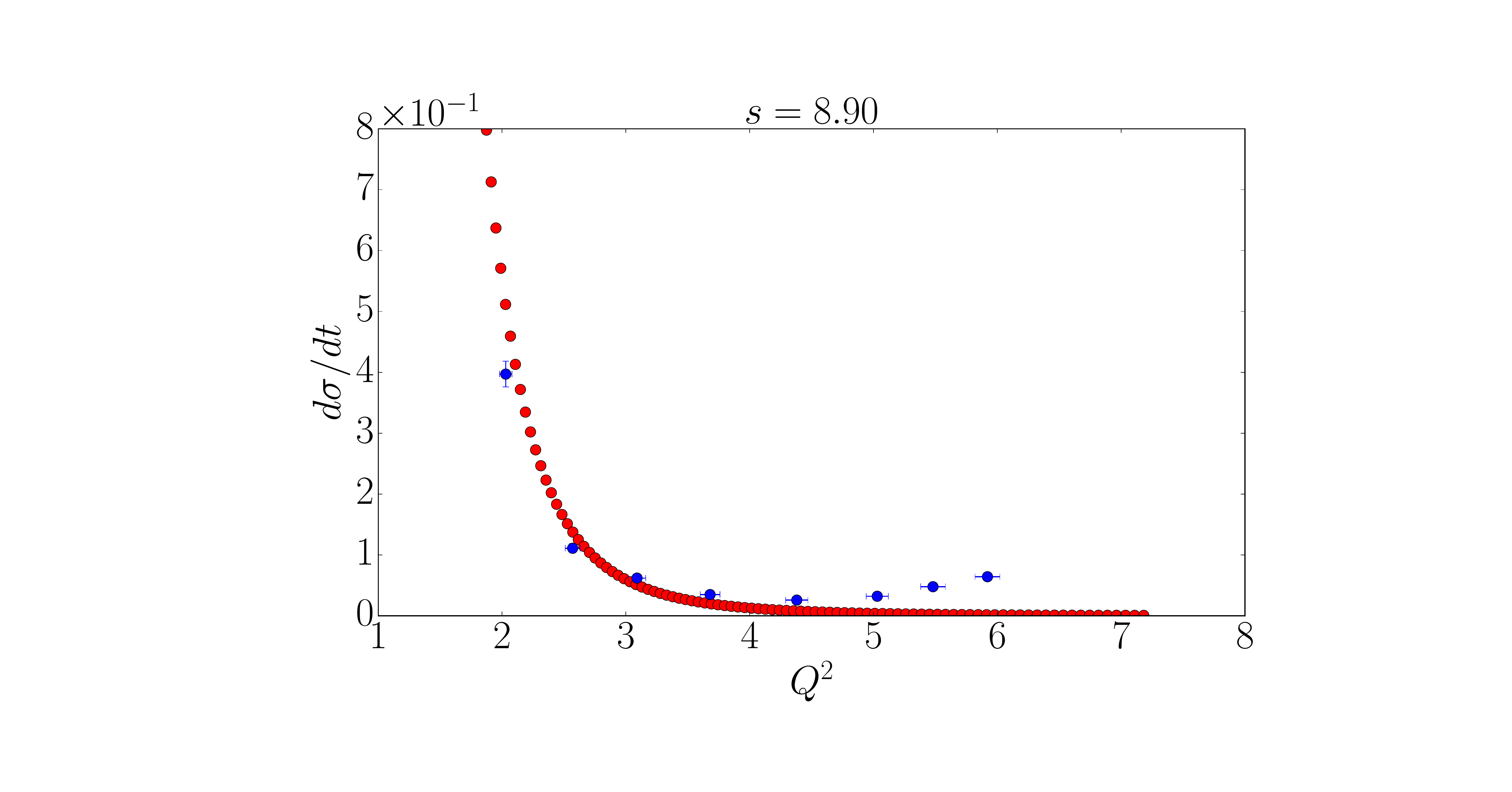}\\
\includegraphics[width=3.8in,keepaspectratio]{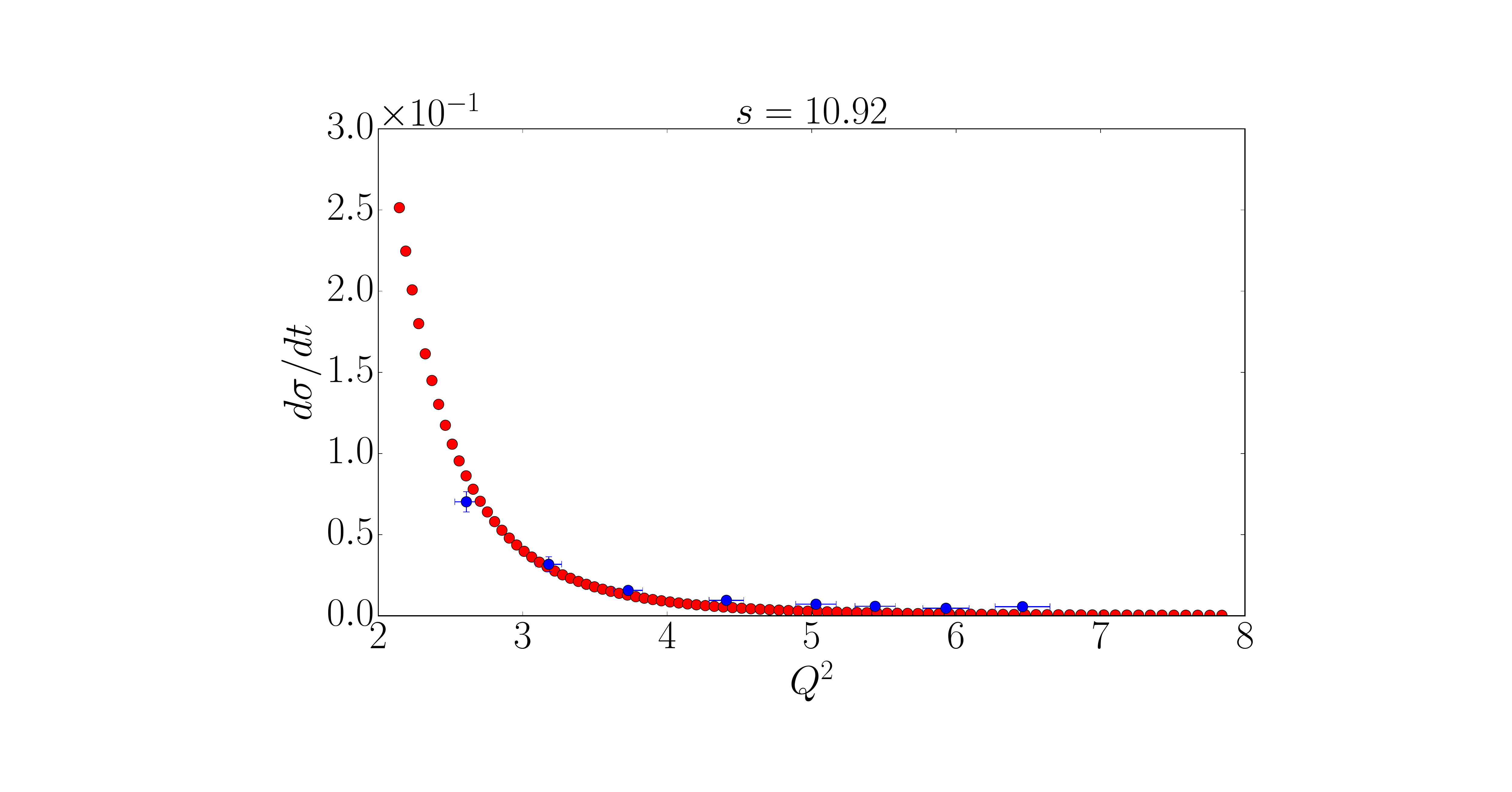}
\caption{Plot of $\frac{d\sigma}{dt}$ nbarns/GeV$^2$ vs  t for $s = 6.79, 8.90,
10.92$ GeV$^2$ and $m=0.29 \mathrm{\, GeV}, \lambda = 1/2, v = -16, a= 0, t = 45$}
\label{fig:fixs}
\end{figure}

At fixed $s$ much larger than $t$, we observed that the experimental data showed a scaling behaviour of 
$1/t^4$ at lower angles. We carried out EP calculations at $s$ = $6.79$, $8.90$, $10.92$ $\mathrm{\, GeV}^2$
and observed that the scaling can be correctly reproduced by the model as was also seen in the
calculation in Sec \ref{sec:scaling}.
We can see in  Fig. \ref{fig:fixs} that there is a good agreement between the data and our EP
prediction at the above energies, which improves as we increase the
$s$ of the data. The rise in the
$d\sigma / dt $ at larger angles is however not captured by the EP calculation. Our choice of $\theta_{CM} =
90^{\circ}$ in the fixed $\theta_{CM}$ analysis above was influenced by this disagreement.

\begin{figure}[]
\centering
\includegraphics[width=3.8in,keepaspectratio]{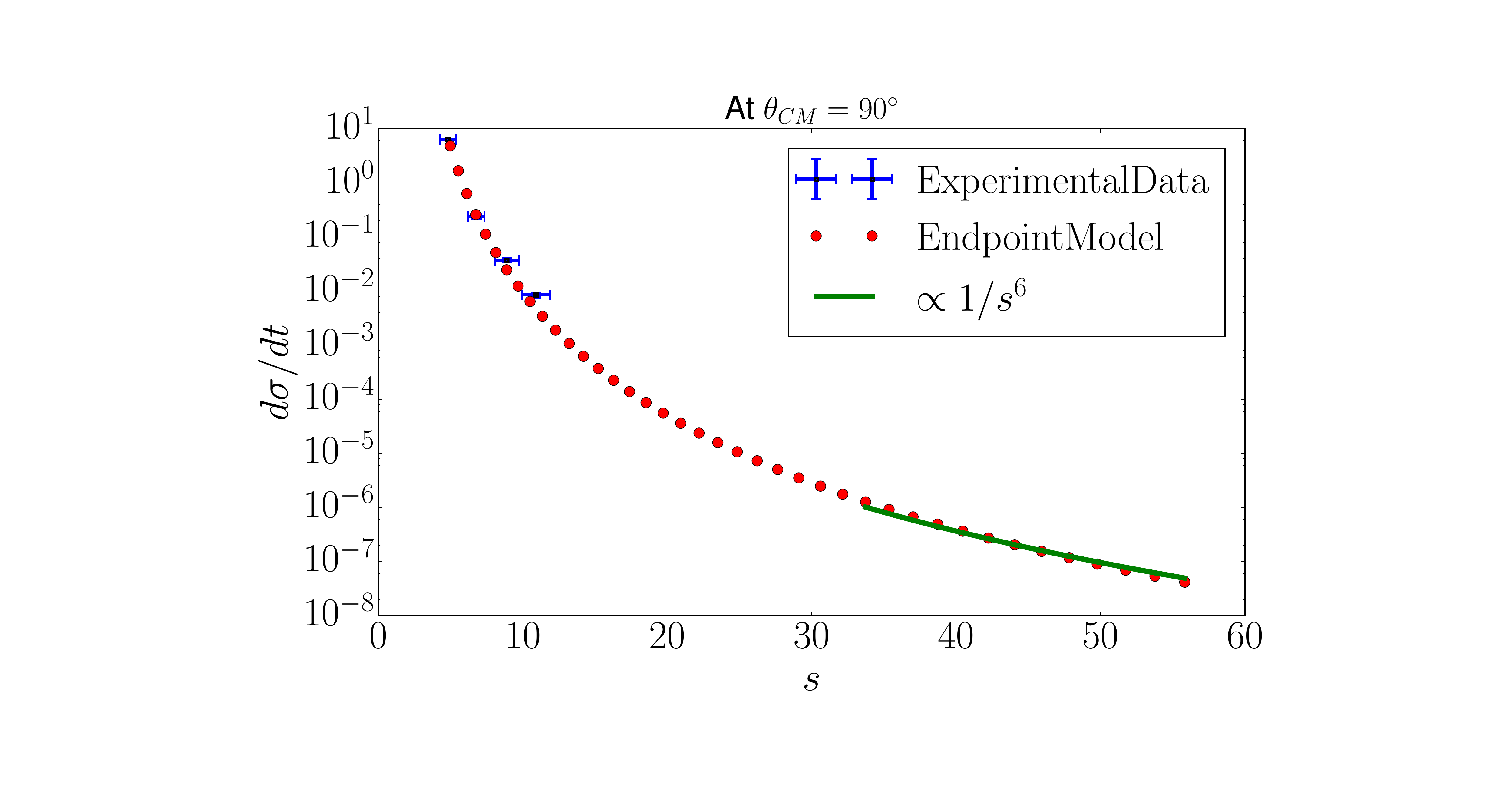}
\caption{EP evaluation of $d\sigma/dt$ $\frac{\mathrm{nbarns}}{\mathrm{GeV}^2}$ vs $s$ GeV$^2$
for $m=0.29 \mathrm{\, GeV}, \lambda = 1/2, v = -16, a= 0, t = 45$}
\label{fig:s6}
\end{figure}

For the fixed $\theta_{cm}$ analysis in the $s \sim t$ region, the expected scaling behaviour from the quark counting rules \cite{Brodsky:1973kr,Matveev:1973ra}
was $1/s^6$ and
is not seen in the experimental data which shows a scaling of $1/s^8$\cite{Danagoulian:2007gs}
. We evaluate the integral in Eq. \ref{eq:dsdt} for a range of
$Q^2$ at $\theta_{CM} = 90^{\circ}$ and we observe in the resulting plot (Fig. \ref{fig:s6}) that EP
shows the above  scaling behaviour of $1/s^6$ after we reach $s\sim 25 \mathrm{\, GeV}^2$.  
At the experimental energy levels, though a $1/s^8$ scaling was not observed, there was a remarkable match
between the EP predictions and the experimental data.

\section{Conclusions}

The Endpoint model combines the soft mechanism and the nature of the transverse momenta of a quark
in a hadron to study scaling behaviour in its exclusive processes. Using the model to calculate
exclusive
processes leads to expressions dominated by the endpoint region of the wavefunction, this helps us
extract the nature of the wavefunction. Specifically for the proton, using one set of data to obtain
the wavefunction ($F_1$ data), the scaling behaviour of $F_2/F_1$ of proton and $pp$ scattering was
successfully obtained. The successes of the Endpoint model lead us to the problem of real Compton
scattering of the proton.

The experimental data for Compton scattering \cite{Danagoulian:2007gs}
show a scaling behaviour for the differential
scattering cross section in two regions of $s,t$: a $1/s^8$ scaling for fixed $\theta_{CM}$ and $s \sim t \gg \Lambda_{QCD}$ 
and a $1/t^4$ scaling at fixed $s$ much larger than $t$. 
Fixing the free parameters in the Endpoint Model using the data for  $F_1$ and $F_2/F_1$, we carried
out numerical calculation for Compton scattering in these limits.
For fixed $s$ larger than $t$, the Endpoint calculations show the $1/t^4$ scaling observed in data
and have a good match with the data for lower angles. 
In the fixed $\theta_{CM}$ and $s \sim t \gg \Lambda_{QCD}$ region, the Endpoint Model calculation for the
Compton scattering shows the elusive $1/s^6$ scaling, that is expected from constituent
counting rules \cite{Brodsky:1973kr,Matveev:1973ra}. Moreover, the Endpoint model also suggests that the $1/s^6$ scaling can be expected
to be dominant after $s \sim 25 \, \mathrm{GeV}^2$.
At the experimental values of $s$, though the experimentally observed scaling is absent in the
Endpoint Model, an excellent
agreement with experimental observations can be seen when extending the calculation to lower $s$
(lower
\,$Q^2$).

With the current work, we have shown once again that the Endpoint model is capable of explaining a range
of scaling laws for hadronic processes. It is capable of generating the quark counting
rules \cite{Brodsky:1973kr,Matveev:1973ra} and also suggests the energy scales at which one can
expect these
scaling laws to dominate. Fixing the parameters of the model using existing data, the Endpoint model
is also able to give an excellent match with experimental data.

As we go to higher angles in the fixed $s$ differential cross section measurements, EP does not
correctly predict the rise in the $d\sigma / dt$ which has to be explored in future work.
Also, further work will be required for the  evaluation of polarization transfer variables ($K_{LL} \&
K_{LS}$) under the Endpoint model.

\section*{Acknowledgement}
The author would like to thank Prof. Pankaj Jain for useful discussions and comments. For the
computational work in this paper, I would like to thank the Physics Department at IIT Kanpur for
facilities provided. The author would also like to thank Bogdan Wojtsekhowski for suggesting the
Compton scattering problem.

\end{document}